%% file: author_v3.tex
\documentclass[graybox]{svmult}

\usepackage{natbib}
\usepackage{mathptmx}       
\usepackage{helvet}         
\usepackage{courier}        
\usepackage{type1cm}        
\usepackage{makeidx}         
\usepackage{graphicx}        
\usepackage{multicol}        
\usepackage[bottom]{footmisc}
\usepackage{url}

\makeindex             

\begin{document}

\title*{ARES+MOOG - a practical overview of an EW method to derive stellar parameters}
\author{S\'ergio G. Sousa}
\institute{S\'ergio G. Sousa \at Centro de Astrof\'isica, Universidade do Porto, Rua das 
Estrelas, 4150-762 Porto and Departamento de F\'isica e Astronomia, Faculdade de Ci\^encias, 
Universidade do Porto, Rua do Campo Alegre, 4169-007 Porto, Portugal, \email{sousasag@astro.up.pt}}
%
%

\titlerunning{ARES+MOOG: a practical overview of an EW method}
\maketitle

\abstract*{The goal of this document is to describe the important practical aspects in the use of an Equivalent 
Width (EW) method for the derivation of spectroscopic stellar parameters. A general description 
of the fundamental steps composing any EW method is given, together with possible 
differences that may be found in different methods used in the literature. Then ARES+MOOG is then used as an 
example where each step of the method is described in detail. A special focus 
is given for the specific steps of this method, namely the use of a differential analysis to define the atomic 
data for the adopted line list, the automatic EW determinations, and the way to find the best parameters at the end of the procedure.
Finally, a practical tutorial is given, where we focus on simple exercises useful to illustrate and explain the dependence 
of the abundances with the assumed stellar parameters. The interdependences are described and a clear 
procedure is given to find the ``final'' stellar parameters.
}

\abstract{The goal of this document is to describe the important practical aspects in the use of an Equivalent 
Width (EW) method for the derivation of spectroscopic stellar parameters. A general description 
of the fundamental steps composing any EW method is given, together with possible 
differences that may be found in different methods used in the literature. Then ARES+MOOG is then used as an 
example where each step of the method is described in detail. A special focus 
is given for the specific steps of this method, namely the use of a differential analysis to define the atomic 
data for the adopted line list, the automatic EW determinations, and the way to find the best parameters at the end of the procedure.
Finally, a practical tutorial is given, where we focus on simple exercises useful to illustrate and explain the dependence 
of the abundances with the assumed stellar parameters. The interdependences are described and a clear 
procedure is given to find the ``final'' stellar parameters.
}

\section{Introduction}
\label{sec:1}

For the derivation of spectroscopic stellar parameters people normally choose between two 
possible methods. One is normally referred as the ``spectral synthesis method'', the other is referred as the 
``Equivalent Width (EW) method''. The spectral synthesis method typically starts with the synthesis of 
theoretical spectra which are then compared to the observed spectrum. In this case the ``final'' 
parameters are found when the correspondent synthetic spectrum fits the observational spectrum. 
Alternatively, the EW method starts directly with the observed spectrum, 
measuring the strength of selected and well-defined absorption lines which are translated into 
individual line abundances, assuming a given atmospheric model. Then, a comparison between the 
computed abundances and the respective theoretical predictions is performed in order to find 
the ``final'' parameters. 

It is clear that both approaches have their own advantages and disadvantages. From one side, the EW method 
can be faster than the synthetic method since it is focused on only a specific number of lines, while the 
synthetic method needs a more complete description of the spectrum. On the other hand, if the individual 
lines used by the EW method cannot be properly isolated then this may lead to inaccurate results. 

The goal of this document is to give a description of the ``ARES+MOOG method''
which is based on the EWs. ARES is the code for automatic EW measurements of the observed 
spectrum \citep[see][]{Sousa-2007}, and MOOG is used to perform the 
individual abundance calculations \citep[see][]{Sneden-1973}. The method ARES+MOOG, like other 
EW methods, allows us to derive the stellar atmospheric parameters: effective temperature ($T_{\rm eff}$), 
surface gravity ($\log{g}$), microturbulence ($\xi$), and the iron abundance ([Fe/H]). The method makes 
use of the excitation and ionization balance from the iron lines where the [Fe/H] is used as a proxy 
for metallicity. This method has been successfully applied to several large samples of F, G, and K 
(FGK) solar type stars \citep[see e.g.][]{Sousa-2008,Sousa-2011b}. 

\section{EW Method: a General Overview}
\label{sec:2}

\begin{figure}[t]
\includegraphics[scale=.27]{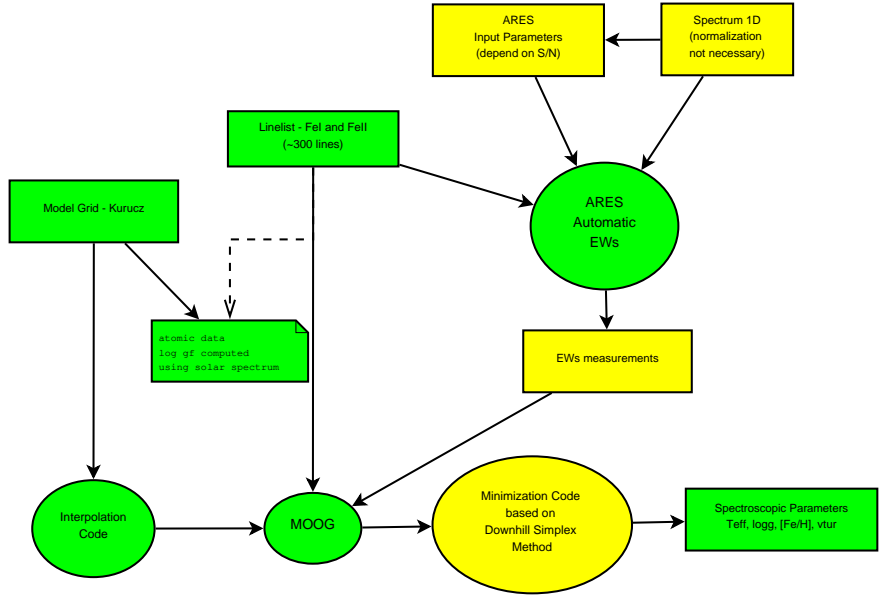}
%
%
\caption{The workflow diagram of the ARES+MOOG method.}
\label{fig:1}       
\end{figure}

Figure \ref{fig:1} reports the workflow describing, as an example of the EW method, the ARES+MOOG procedure 
to derive the stellar atmospheric parameters. From this diagram we can easily identify the general steps for an EW 
method based on the excitation and ionization balance of iron lines:

\begin{enumerate}
\item A list of iron absorption lines with the correspondent atomic data is selected for the analysis;
\item The observational spectrum is analysed and the EWs are measured independently in a line-by-line analysis;
\item A stellar atmospheric model is adopted given the atmospheric parameters;
\item The measured EWs and the atmospheric models are used to compute the individual line abundances;
\item The ``final'' spectroscopic parameters are found once the excitation and ionization balance is achieved for all the individual lines 
	analyzed, otherwise we go back to step 3 and adopt different parameters;
\end{enumerate}

These are the basic steps required for the use of an EW method. The differences between the 
EW methods found in the literature are typically centered on the use of different line lists, models 
and codes used in each step.

%

\subsection{Line-list}

The selection of the lines to be used in the analysis (first step) is crucial for the accuracy and precision 
of this method. Some authors use a large set of lines with the aim to increase 
the statistical strength of the derived spectroscopic stellar parameters. Others authors 
use a reduced and very well-defined set of absorption lines which are considered to be well 
known or, at least, very well adapted for a specific type of stars (e.g. giant stars).

Together with the line selection, the adopted atomic parameters for each line are also of paramount 
importance. Although we can find very accurate wavelengths and excitation 
potentials for each line, the oscillator strengths ($\log gf$) are not so precisely known. The 
uncertainties of these values, which can be measured in laboratory, can propagate 
and affect dramatically the precision and accuracy of the derived spectroscopic 
parameters.

\subsection{Measurement of EWs}

Although the definition of the an EW is quite simple, its measurement from an observed spectra can 
still be tricky (second step). The determination of the correct position of the continuum level 
continues to be an important source for the uncertainties in these measurements. Another important 
aspect is to understand if a given line that needs to be measured is completely isolated (the 
ideal case) or if there are close-by lines that are blended together. In the latter  
case the correct identification of all lines is fundamental for a good measurement.

Moreover, for the EW measurements is important to define the profile function used for the line fit and 
corresponding strength calculation of each lines. The Gaussian profile is widely used and is considered to 
be an almost perfect approximation for weak absorption lines. However some caution should be taken when 
measuring strong lines (typically EW $>$ 150 m$\AA$). In this case the Gaussian profile cannot 
perfectly fit the wings of the line. In these cases several authors prefer to use the Lorentzian profile.

Until recently, these measurements were only feasible using interactive routines such as the ``splot'' task in IRAF\footnote{IRAF is 
distributed by National Optical Astronomy Observatories, operated by the Association of Universities for Research in Astronomy,
Inc., under contract with the National Science Foundation, USA}. In this case, people have to go through the spectrum line by line, 
marking the continuum position by eye and make the EW measurement. 
This is, of course, a quite boring and very time consuming task. Even worse, the subjectivity involved in such interactive routines may 
lead to inconsistency between the measurements of different lines. To overcome this issues, several 
automatic codes are now on the market (e.g. ARES, \citet[][]{Sousa-2007}; DAOSPEC, \citet[][]{Stetson-2008}) that measure the EWs in a more efficient
and consistent way.

\subsection{Model Atmospheres}

The literature offers the possibility to choose from a wide variety of model atmospheres (third step) 
and often this choice can be affected by subjectivity. Models like ATLAS9 \citep[][]{Kurucz-1993} and 
MARCS \citep[][]{Gustafsson-2008} standout as the most used set of atmospheric models 
for the derivation of spectroscopic and elemental abundances. 

The way the models are created and used in each method can also differ. Some authors prefer to create the models on the spot making use
of available codes. As an alternative, and in order to increase the efficiency of each method, grids are created with pre-computed 
model atmospheres, which can be directly selected for each step of the iteration, or instead, the grid of models can be 
used for interpolations allowing this way a better refinement in the search of the ``final'' stellar parameters.

As one could expect, for the creation of the models there are a series of important physical 
parameters and approximations that need to be defined and used for the correspondent computation. For instance, for 
FGK solar-type stars the plane-parallel approximation has been proved to 
be a safe approach, but specific models may be necessary when dealing with ``special'' types of stars (e.g., metal-poor stars, 
giant and evolved stars).

\subsection{Computing Abundances}

In step number four we have the computation of the iron abundance (or any other chemical element). This step clearly depends 
on the measured strength of the line as well as the selection of the atmospheric model. 

As for such models, local thermodynamical equilibrium (LTE) is commonly assumed as a valid approximation for FGK solar-type 
stars. However, this approximation may not be completely valid for other types of stars (e.g., for very metal-poor 
stars NLTE corrections may be necessary to apply \citep[][]{Bergemann-2012}.

\subsection{Finding the correct parameters}

In step five we need to achieve the excitation and ionization equilibria. The correlation between the excitation potential 
and the iron abundance of each line constrains the effective temperature, the correlation between the reduced 
equivalent width and the individual abundance constrains the microturbulence, and the ionization 
balance between the mean abundance of FeI and FeII fixes the surface gravity. The "final" stellar parameters will be obtained 
when no correlations are present, i.e. when all lines give the same individual abundances we stop the process and keep the 
parameters from the adopted model atmosphere. The iron abundance comes as an additional 
result from this analysis and is taken as the mean abundance from all lines.

The main difference between the various EW methods in this step may be related with the way in which the parameters are found and constrained. There are 
different minimization algorithms that may be used to explore the parameter space and the respective inter-dependences. Finally the 
constraints used to stop the method and check for the correct convergence of the parameters (i.e. what is the definition of no 
correlation from the data, e.g. does a slope with a value of 0.01 represents 
no correlation?) are crucial for the final decision of each method that may change the resulting parameters.

\section{ARES+MOOG: the method}
\label{sec:3}
So far it was described a general overview of an EW method. Here we will go through the steps 
once again and describe the specific choices made in the ARES+MOOG (whose workflow is described in Fig. \ref{fig:1}).

\subsection{Line list}

Since this method was designed to be completely automatic, for the compilation of the line list we have selected as many 
lines as possible. This increases the statistical strength of the derived parameters. However, each line was 
carefully selected in order to be considered stable for this method \citep[for further details on the stability of the 
lines see][]{Sousa-2008}.

Regarding the atomic data, in order to overcome the uncertainties for the $\log gf$ described before, we made use of a 
differential analysis technique. This technique consists in selecting a benchmark star (typically the Sun) with very 
well constrained parameters. The goal of this analysis is to recompute the $\log gf$ using an inverse analysis. We first 
measured EWs for our selected lines in the solar spectrum. Then, assuming the solar parameters (e.g. $T_{\rm eff}$ = 5777K, $\log g$ = 4.44 dex, 
$\xi_{\mathrm{t}}$ = 1.00 kms$^{-1}$, and $\log (\mathrm{Fe})$ = 7.47), the values for each $\log gf$ were then changed until we derive the ''correct`` 
individual abundance for each line.

Using this differential analysis it is possible to reduce both the errors on the atomic parameters and the errors 
on measurements of the equivalent widths. When measuring the lines in a benchmark 
star we are also including errors given by its spectrum itself. For instance, the existence of small undetected blended lines or the
intrinsic bad position of the continuum for each line will introduce errors when computing the $\log gf$. The differential analysis will allow us to 
partially compensate for such kind of errors by assuming a systematic measurement of the EW for the same lines 
of the stars analyzed. An obvious drawback from this analysis is that the results strongly deteriorate as we use the $\log gf$ in stars 
that become more and more different from the benchmark star.

\subsection{Measuring EWs}

The Equivalent Width of the lines were automatically determined using ARES\footnote{The ARES code can be downloaded 
at \url{http://www.astro.up.pt/~sousasag/ares/}} code \citep{Sousa-2007} following the approach of \citet{Sousa-2008} and \citet{Sousa-2011a} to adjust 
the $rejt$ parameter of ARES according to the $S/N$ of each spectrum. In the next section 
the input parameters for ARES will be described in detail and some advices will be given in order 
to select the best input parameters.

\subsection{Model Atmospheres}

We used MOOG 2013\footnote{\url{http://www.as.utexas.edu/~chris/moog.html}} \citep{Sneden-1973} to compute the line-by-line abundance 
for each star assuming LTE conditions. In our standard method we used a grid of Kurucz Atlas\,9 plane-parallel model 
atmospheres \citep[][]{Kurucz-1993} in order to generate the appropriate model atmosphere through interpolation. This model is 
then feed as an input into MOOG to compute the abundances through the driver \textit{abfind}.

\subsection{Finding the final parameters and the Iron Abundance}

We use as minimization algorithm to determinate the best stellar parameters the Downhill Simplex Method \citep{Press-1992}. 
Moreover, in order to identify outliers caused by incorrect EW values, we performed a 
3-$\sigma$ clipping for the FeI and FeII lines after a first preliminary determination of the stellar parameters. 
After this, the procedure was executed once again without the rejected lines. For a wider discussion about the full 
automatization of this method see the works of \citet[][]{Santos-2004, Sousa-2011b, Saffe-2011}.

\section{ARES+MOOG: Quick Tutorial}
\label{sec:4}


The tutorial presented here follows the procedures and codes that were made available at 
the ``Spring School of Spectroscopic Data Analyses''. The codes are available either at the 
respective web-pages or are still accessible from the school web-page: \url{http://spectra.astro.uni.wroc.pl/}.

As described before, the first step for the ARES+MOOG method is the definition of the line list. We will use the very well defined line list 
composed of nearly 300 iron lines presented in \citet[][]{Sousa-2008}. If the reader is 
interested in a recent update of the line list see \citet{Tsantaki-2013}.

\subsection{Using ARES}

A complete description of ARES can be found in \citet [][]{Sousa-2007}. In this document we will only point out the essential steps 
required to properly run the code.

A sketch of the ARES procedure is presented in Figure 1 of \citet [][]{Sousa-2007}. The basic steps of ARES are: i) the reading of both the spectrum and 
the line list; ii) the local normalization of the spectrum which is performed for each line 
in each iteration; iii) the detection of the set of lines that are needed to be fitted (in case of blended lines); 
iv) the fit and the measurement of the EWs; v) the storage of the EWs in an output file.


\subsubsection{Preparing the spectrum}

The first step to properly use ARES is the preparation of the observed spectra. The available version of ARES only works with 
one-dimensional FITS spectrum. In the respective FITS header of the spectrum the CDELT1 and CRVAL1 keywords need to be defined as a requirement.

Another fundamental condition is that the spectrum should be corrected in radial velocity so that the absorption lines are found at 
the rest frame, otherwise ARES will not be able to find the correct line position for the analysis.

\subsubsection{The line list}

The only requirement for the line list to be feed in ARES is the correct wavelength. The file with the list 
of lines should consist of a column with the wavelength. ARES will read this file line-by-line for the respective EW measurement.
It may be also useful to keep in this file the atomic data for each line that will be required later on. In Table \ref{tab1} is presented 
a sample of line list from \citep[][]{Sousa-2008}, where for each line is defined the rest wavelength ($\lambda$), the excitation 
potential ($\chi_{l}$), the oscillator strength($\log{gf}$) the element identification (Ele. and Num.) and the EW measured in a solar spectrum (EW$_{\odot}$).

\begin{table}[!ht]
\centering 
\caption[]{Sample of the line-list presented in \citep[][]{Sousa-2008}.}
\begin{tabular}{lccccc}
\hline
\hline
\noalign{\smallskip}
$\lambda$ (\AA) & $\chi_{l}$ & $\log{gf}$ & Ele. & Num. & EW$_{\odot}$\\
\hline
6079.01	&   4.65 &   -1.008 &   FeI  &   26.0 &   45.8  \\
6082.72	&   2.22 &   -3.566 &   FeI  &   26.0 &   34.5  \\
6084.11	&   3.20 &   -3.774 &   FeII &   26.1 &   20.9  \\
6089.57	&   4.58 &   -1.273 &   FeI  &   26.0 &   35.3  \\
...     &   ...  &    ...   &   ...  &   ...  &   ...   \\

\hline
\end{tabular}
\label{tab1}
\end{table}

\subsubsection{ARES input parameters}

The input parameters for ARES are the following:

\begin{itemize}
  \item \textit{specfits}: The name of the 1D fits file with the spectrum corrected in Radial Velocity (e.g. HD1234\_rv.fits).
  \item \textit{readlinedat}: The name of the file with the list of lines to be measured (e.g. line list.dat).
  \item \textit{fileout}: The name of the file that will contain the output of the results (e.g. HD1234.ares).
  \item \textit{lambdai}: Initial wavelength to search the lines (e.g. 3000 \AA).
  \item \textit{lambdaf}: Final wavelength to search the lines (e.g. 7000 \AA).
  \item \textit{smoothder}: Smooth value for the derivatives that are used for the line detection 
			    procedure (e.g. 4 - recommended value for high resolution spectra and good S/N).
  \item \textit{space}: Size of the local spectrum interval in Angstroms around each line. Only 
			  this interval is used for the individual computations of each line (e.g. 3 \AA - recommended value).
  \item \textit{rejt}: Parameter for the calibration of the continuum position. This value strongly depends on the S/N of the spectrum. 
			A good reference for the values to be used here can be found in \citet[][]{Sousa-2008,Sousa-2011a} (e.g. 
			0.996 for spectra with S/N $\sim$ 400).
  \item \textit{lineresol}: This parameter sets the line resolution of the input spectra; This 
			    parameter is helpful to distinguish real lines from noise (e.g. 0.1 \AA - recommended value for 
			    high resolution spectra).
  \item \textit{miniline}: Lines with strength weaker that this value are not printed in the output file (e.g. 2 m\AA).
  \item \textit{plots\_flag}: Flag for the plots (0-runs in batch, 1-shows the plots and stops for each line calculation).
\end{itemize}

There are specific input parameters that are very important to obtain correct EWs. A proper selection of the \textit{rejt} parameter 
is fundamental in order to track the correct continuum position. Wrong values of this parameter may systematically give larger 
(or smaller) EWs. Although there is a clear dependence between this parameter and the S/N we choose to leave this as a free parameter 
given the high degree of subjectivity when defining the continuum position. If some authors want to define their own S/N dependence, 
we advise the reader to select only a few isolated lines for a few spectra with different S/N and to make use of the plots to select 
the best values for each S/N. For more details on such exercise see the work of \citet[][]{Mortier-2013}. For the other parameters, 
the recommended values should be kept fixed. We may only consider to tweak the \textit{smoothder} parameter at 
higher values in case of very low S/N spectra. This may help for the correct identification of real lines in noisy spectra.

\subsection{Generating a model atmosphere}

The computations of specific model atmospheres in ARES+MOOG is done by an interpolation code which in turn uses a grid of pre-computed 
ATLAS9 models. The interpolation of models was choosen here for efficiency purposes and consists in of two separated Fortran codes: the first code performs the 
interpolation itself, while the second one accommodates the model in a file with a specific format readable 
by MOOG. A script named ``make\_model.bash''\footnote{This script can be found together with other codes used in the school 
in: \url{http://spectra.astro.uni.wroc.pl/elements/codes_ARESMOOG.tar}\label{fn:repeat}} is provided in order to run both codes directly. The script 
needs as input the astrophysical parameters ($T_{\rm eff}$, $\log{g}$, $\xi$, and [Fe/H]) to generate the model which will be stored in a file named ``out.atm''.

\subsection{Using MOOG}

MOOG is a code that performs a variety of LTE line analysis and spectrum synthesis tasks. The typical use of MOOG is to 
assist in the determination of the chemical composition of a star. In our case we want to measure individual iron line 
abundances to derive the stellar parameters.

There are several drivers available to run MOOG for several different purposes. The MOOG user's manual has a complete description 
of the several drivers. For ARES+MOOG we make use of the \textit{abfind} driver.

One of the chief assets of MOOG is its ability to do on-line graphics. However in ARES+MOOG the graphics are not used at all. The 
visualization of different plots is quite useful to see the dependences of the different parameters with the individual abundance 
determination. Together with a modified MOOG version (where the internal plots were ignored since it requires a proprietary library),  
it was provided a simple Python code to perform the plots (named read\_moog\_plot.py)\footref{fn:repeat}. This code is used to illustrate the 
parameters dependences and respective correlations (see the next sections).

Another important input for MOOG is the list of atomic data for each line in order to perform individual 
abundance calculations. For this purpose, an additional script was provided (make\_linelist\_local.bash)\footref{fn:repeat}; this script reads the 
output of ARES and the initial line list to create the required format file for MOOG. 

\subsection{Search for the correct model}

For this tutorial we will make use of the solar type star HD1461 for which an HARPS-S@La Silla spectrum with high resolution 
and high S/N was analysed by \citet[][]{Sousa-2008}. The final parameters derived for this star and obtained by the authors with ARES+MOOG are: 
$T_{\rm eff} = 5765 \pm 18$ K, $\log{g} = 4.38 \pm 0.03$ dex, a $\xi = 0.97 \pm 0.02$ Km/s, and $[Fe/H] = 0.19 \pm 0.01$ dex\footnote{For 
a proper description on the estimation of the errors with the ARES+MOOG method, see \citet[][]{Sousa-2011a}}.

\begin{figure}[t]
\begin{center}
\includegraphics[width=2.25in]{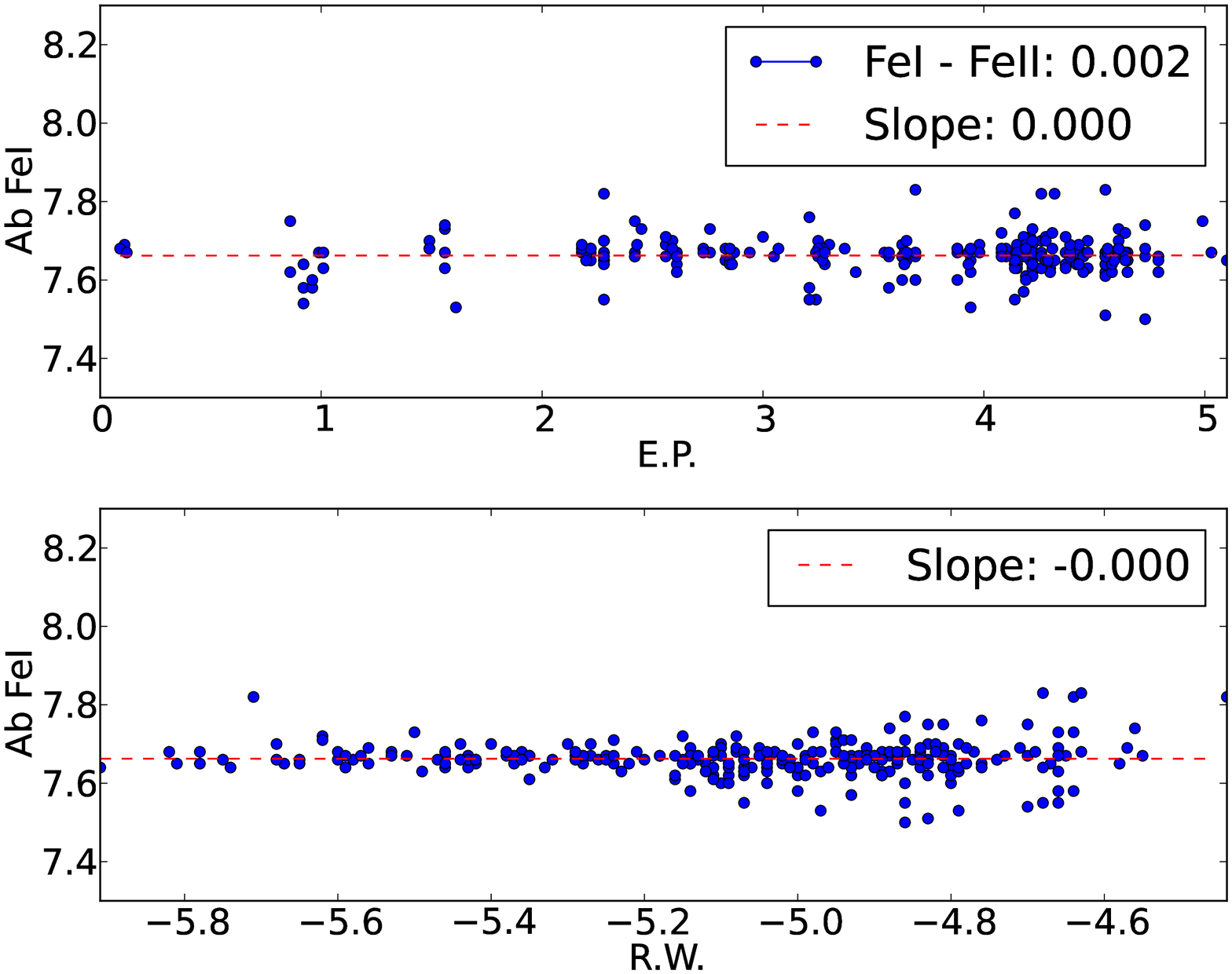}
$\begin{array}{cc}
\includegraphics[width=2.25in]{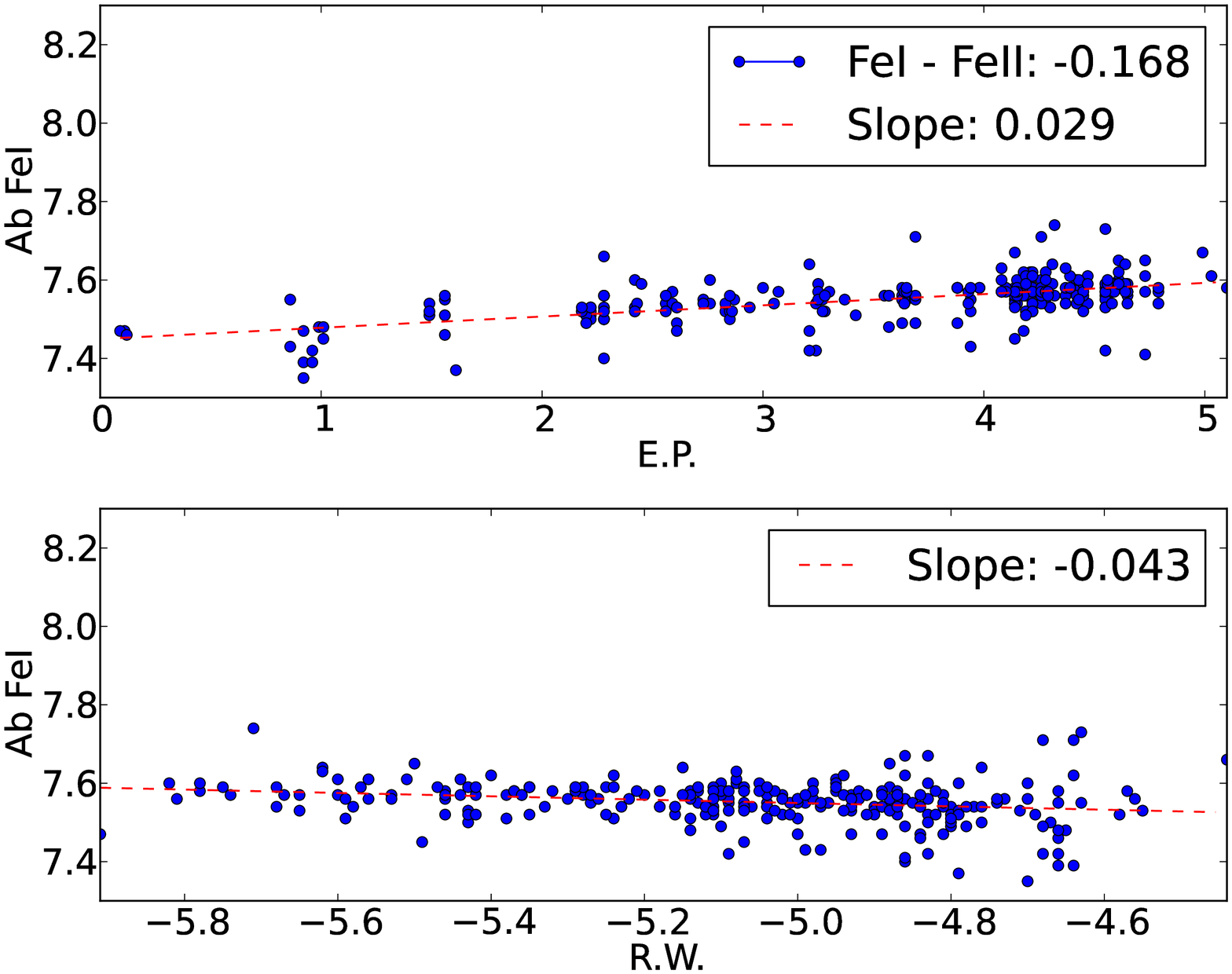} & \includegraphics[width=2.25in]{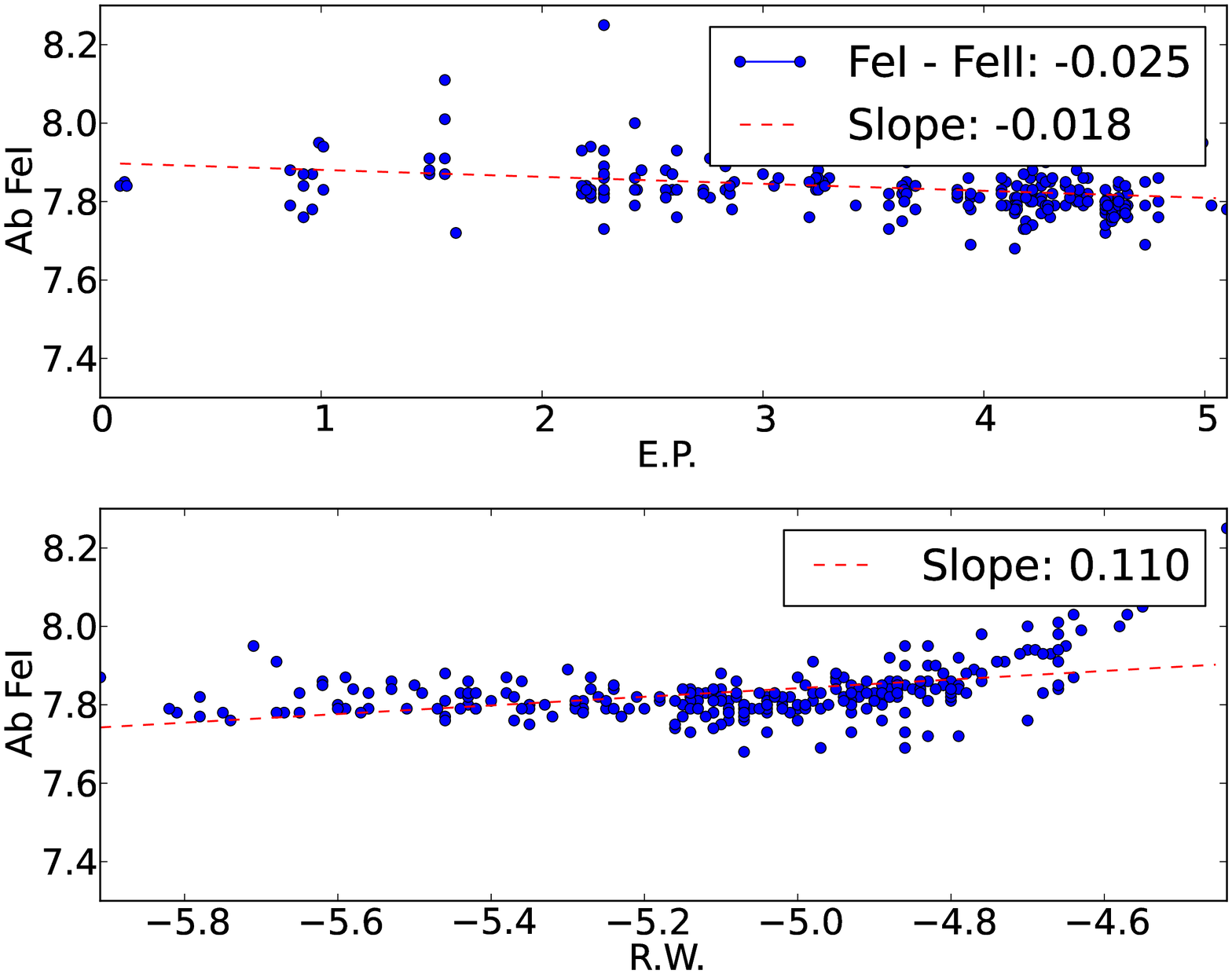} \\
\end{array}$
\end{center}
\caption{Abundance of FeI as a function of excitation potential(E.P.) and reduced wavelength (R.W.). The top panel shows the result for 
the ``final'' stellar parameters, while in the bottom panels the temperature was changed to a lower value (5600 K; left panel), and 
to an upper value (5900 K; right panel).}
\label{fig:3}
\end{figure}

Fig. \ref{fig:3} shows the correlations between the iron abundance (Ab(FeI)) and the excitation 
potential (E.P.) and the reduced equivalent width (R.W.) for all the used iron lines of HD1461. In the same figure it is also indicated together with the 
respective slopes of the correlations, the difference between the average abundances of FeI and FeII ($<$Ab(FeI)$>-<$Ab(FeII)$>$). From the 
values indicated in the figure we can see that the slopes of the correlations obtained with the "final" stellar parameters are close to zero, as 
well as the difference between FeI and FeII.

From theoretical studies, it is possible to demonstrate the the $T_{\rm eff}$ has a strong influence in the 
correlation Ab(FeI) vs. E.P., the microturbulence in the correlation AB(FeI) vs. R.W., and the surface gravity 
is connected directly with $<$Ab(FeI)$>-<$Ab(FeII)$>$ \citep[see, e.g.][]{Gray-2005}. Therefore, here in the following, 
we will make a series of exercises with the aim to show the dependence of each correlation with the spectroscopic 
parameters, and to illustrate how the ``final'' stellar parameters were derived for HD1461. In particular, we will 
show these dependences in a practical way making use of the codes provided for the ARES+MOOG method.

\subsubsection{Effective temperature dependence}

The lower panels of Fig. \ref{fig:3} show the computed abundances for a model with exactly the same parameters as the final ones with exception of 
the temperature. It is clear that the slopes of the correlations change dramatically. Not only the Ab(FeI) vs. E.P. changes but also 
the same happens for Ab(FeI) vs. R.W. showing that the stellar parameters are strongly inter-dependent. 

Hence, these plots show how Ab(FeI) vs. E.P. varies with the changes in temperature so we can react accordingly to find 
the correct temperature. In particular, when we underestimate the temperature the slope is positive, while when we overestimate the 
real temperature, the slope becomes negative. Therefore, the slope gives us information about the direction where the correct temperature is.

\subsubsection{Surface Gravity dependence}

\begin{figure}[t]
\begin{center}
$\begin{array}{cc}
\includegraphics[width=2.25in]{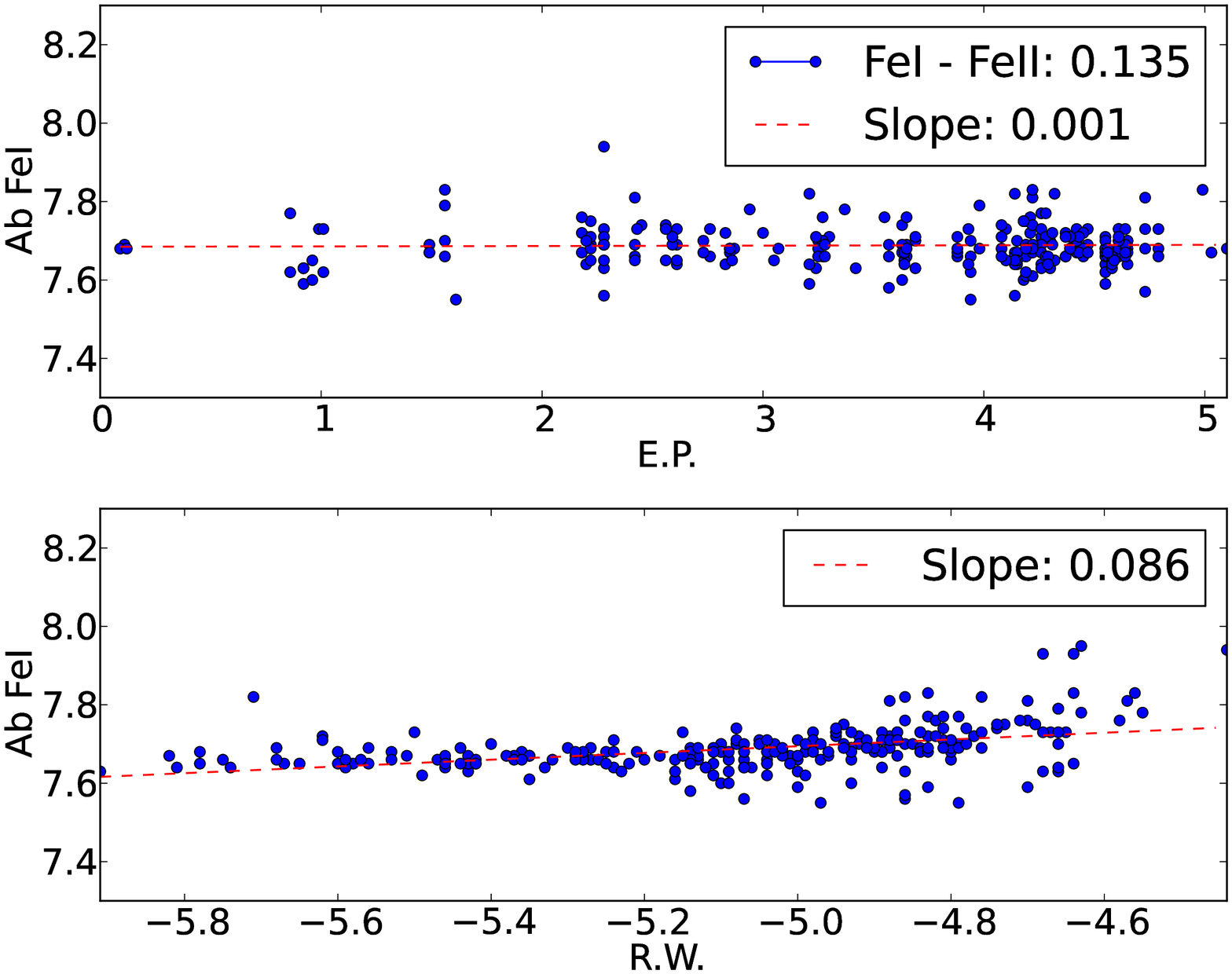} & \includegraphics[width=2.25in]{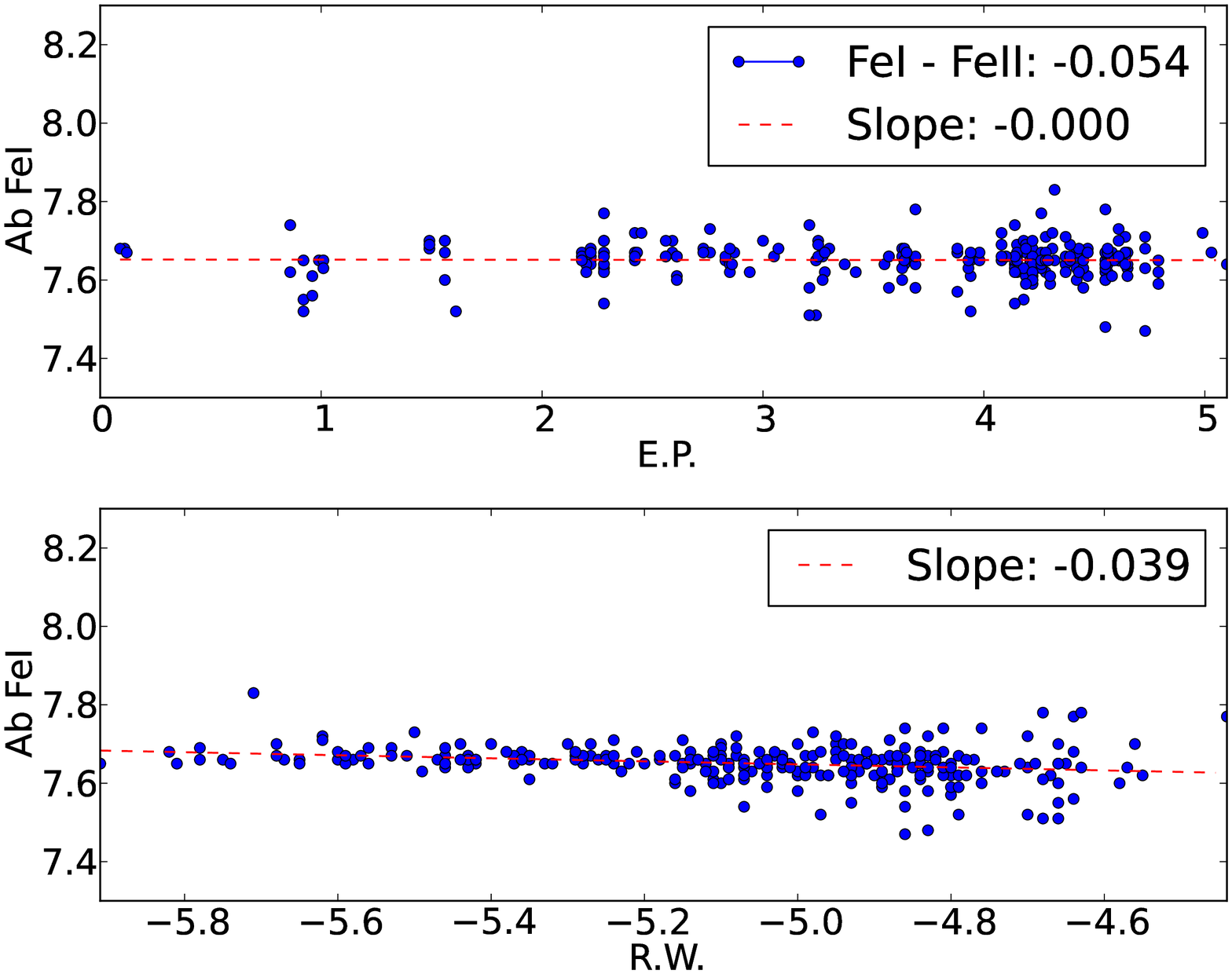} \\
\end{array}$
\end{center}
\caption{Same as Fig. \ref{fig:3} but instead of changing the temperature, here we changed 
the surface gravity to a lower value (4.10 dex; left panel), and an upper value (4.50 dex; right panel).}
\label{fig:4}
\end{figure}

A similar exercise can be done for the surface gravity. Here the temperature was set back to its ``final'' value 
and only the gravity was changed to observe how $<$Ab(FeI)$>-<$Ab(FeII)$>$ varies accordingly. Figure \ref{fig:4} 
shows the results. In particular, when we underestimate the surface gravity $<$Ab(FeI)$>-<$Ab(FeII)$>$ is positive, 
while when we underestimate the surface gravity the $<$Ab(FeI)$>-<$Ab(FeII)$>$ is negative.


One interesting fact is that the changes on the surface gravity nearly does not affect Ab(FeI) vs. E.P. (see 
Fig. \ref{fig:4}). This means that the $\log g$ derived through this method is almost independent on the 
temperature, and vice-versa. This is certainly an advantage 
of this method showing that the temperature and the iron abundance are independently well constrained. A clear disadvantage 
here is that the $\log g$ is not very well constrained, due to the reduced number of ionized iron lines compared to FeI lines. 
This means that extra caution should be considered for the derived values of the surface gravity. For more details on this issue 
see the work of \citet[][]{Torres-2012}.

\subsubsection{Microturbulence dependence}

\begin{figure}[t]
\begin{center}
$\begin{array}{cc}
\includegraphics[width=2.25in]{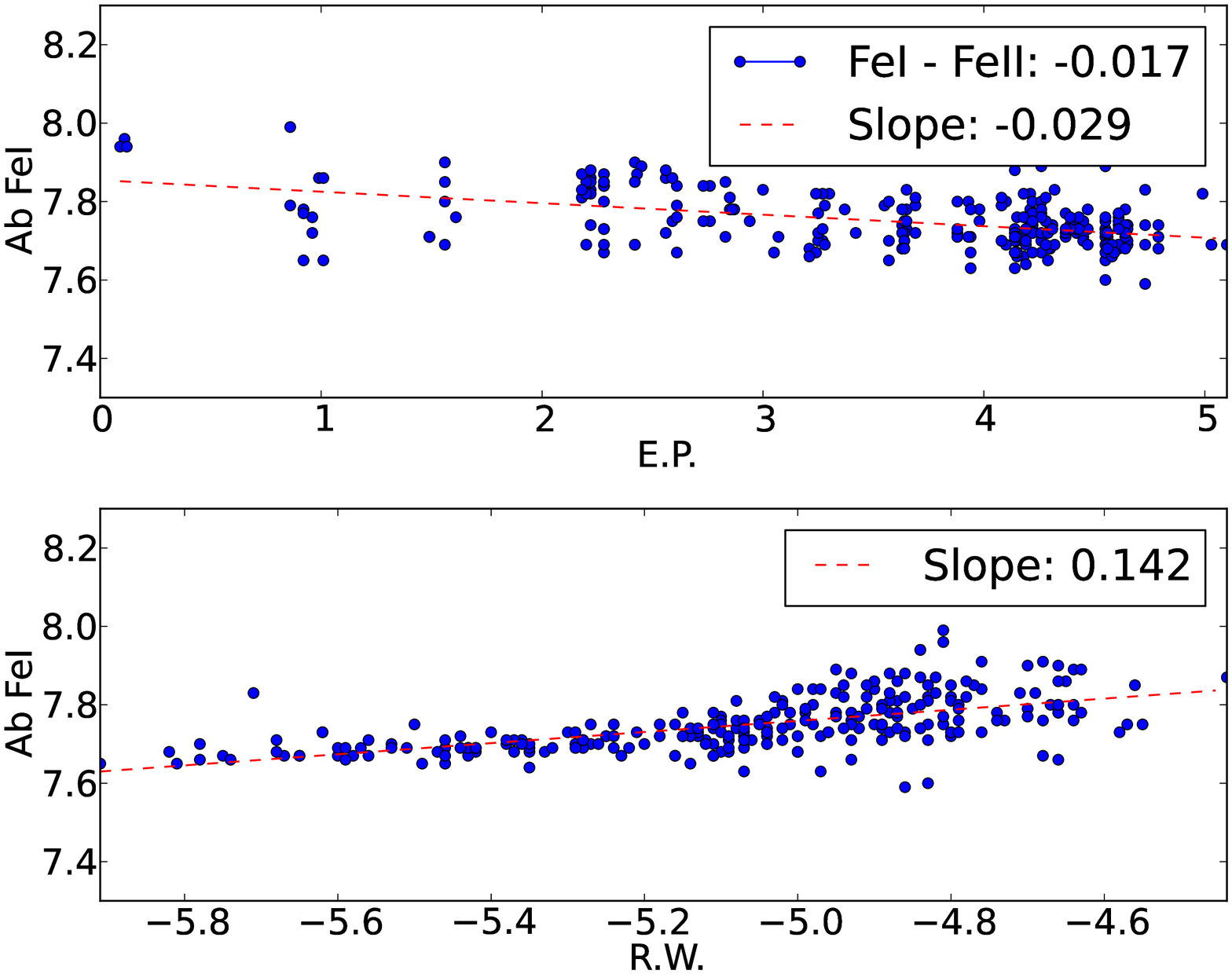} & \includegraphics[width=2.25in]{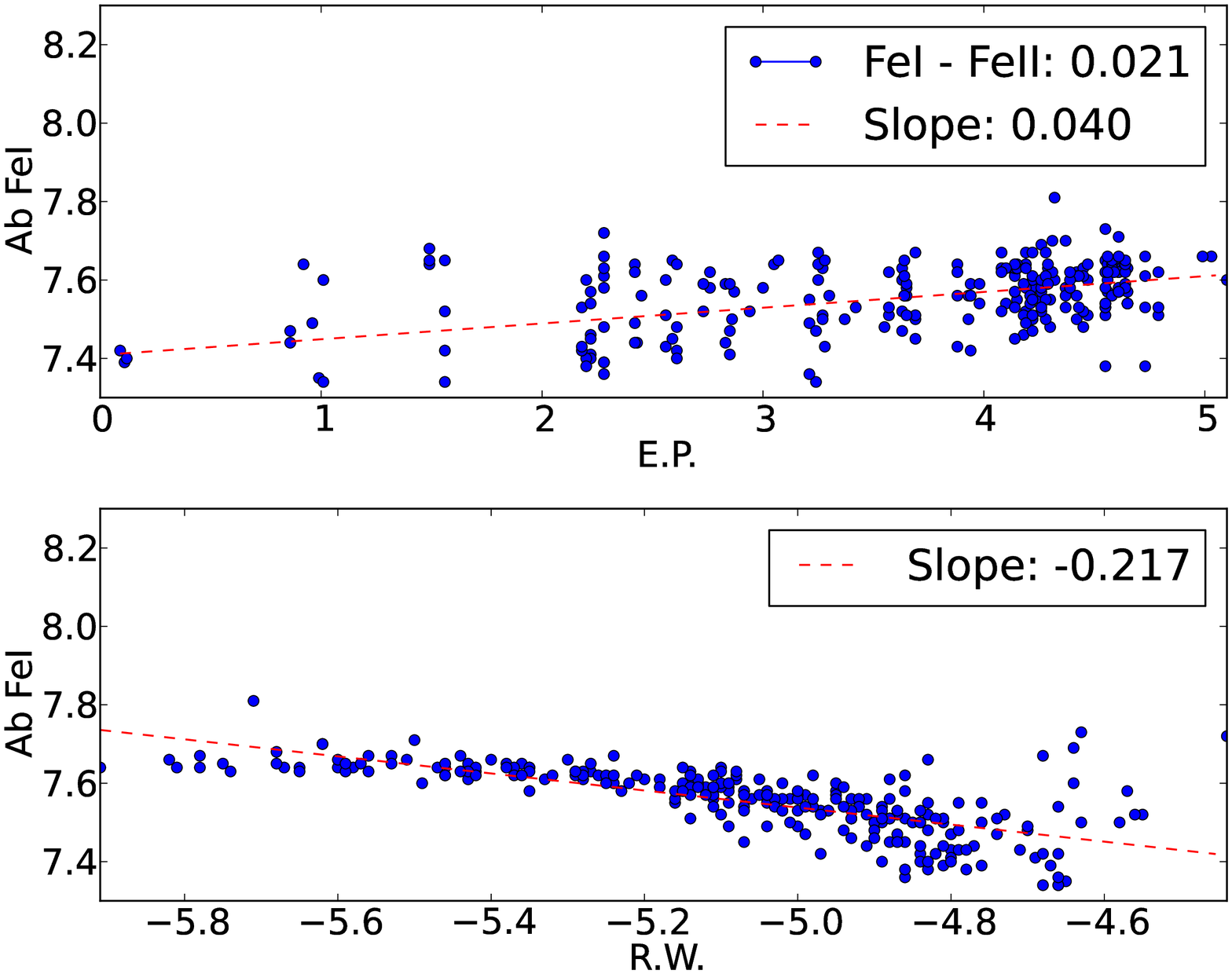} \\
\end{array}$
\end{center}
\caption{Same as Fig. \ref{fig:3} but instead of changing temperature, here we change 
the microturbulence to a lower value (0.5 km/s; left panel), and an upper value (1.5 km/s; right panel) than the final $\xi$.}
\label{fig:5}
\end{figure}

A final exercise can be made for the microturbulence. Again all parameters are set to the ``final'' values with the 
exception of the adopted microturbulence. This parameter is connected with the saturation of the stronger iron lines.
A good value for the microturbulence will allow us to derive the same abundances for weak and strong iron lines.

The left panel of Fig. \ref{fig:5} shows the result of the abundances when the microturbulence is underestimated. The slope 
of Ab(FeI) vs. R.W. is positive in this case. This means that the ``final'' value for the microturbulence should be higher. The 
opposite happens when we overestimate the microturbulence, as can be seen from the right panel of Fig. \ref{fig:5}.

\subsubsection{Final Remark}

Finally, an additional detail regarding the model atmosphere should be considered. When the final parameters are derived, the 
resulting iron abundance (derived from the average of the measured line equivalent widths) must be compatible with the metallicity 
of the model atmosphere.

\section{Summary}
\label{sec:5}

In this document we described in practical terms the use of the EW method to derive spectroscopic stellar parameters. We  
made a general overview of the several steps required to use this method. We described several options used 
by different authors, namely the use of different line lists, and model atmospheres. 

The ARES+MOOG method was described in some detail where we tried to give the best advices for a proper use of it, especially in what 
regards the use of the ARES code to automatically compute the equivalent widths.

The details on how the method finds the ``final'' set of stellar parameters are exposed here. From the practical 
point of view, the essential steps of the method are described and can be used as a guideline for future works. Some additional points 
to fully complete the description of ARES+MOOG were left a side. These include the minimization algorithm which allows a proper automatization 
of the full process and the estimation of the uncertainties.
\begin{acknowledgement}
S.G.S acknowledges the support from the Funda\c{c}\~ao para a Ci\^encia e Tecnologia (Portugal) and FSE/POPH in the form of 
the grants SFRH/BPD/47611/2008 and the scientific cooperation project FCT/Poland 2011/2012 (Proc. 441.00 Poland). S.G.S also acknowledges 
the support by the European Research Council/European Community under the FP7 through a Starting Grant (ERC-2009-StG-239953).
\end{acknowledgement}
\include{referenc}
\end{document}

%% file: referenc.tex
%
%
%
\newcommand*\aap{A\&A}

%% file: author_v3.bbl
\begin{thebibliography}{}
\bibitem[Bergemann et al. (2012)]{Bergemann-2012} Bergemann, M.; Lind, K.; Collet, R., et al.\ 2012, MNRAS, 427, 1, 27
\bibitem[Gray, David F. (2005)]{Gray-2005} Gray, David F. 1995, The Observation and Analysis of Stellar Photospheres, 3rd Edition
\bibitem[Gustafsson et al. (2008)]{Gustafsson-2008} Gustafsson, B., Edvardsson, B., Eriksson, K., et al.\ 2008, \aap, 486, 951
\bibitem[Kurucz et al. (1993)]{Kurucz-1993} Kurucz, R. 1993, ATLAS9 Stellar Atmosphere Programs and 2 km/s grid. Kurucz CD-ROM No. 13. Cambridge, Mass.: Smithsonian Astrophysical Observatory, 1993., 13
\bibitem[Mortier et al. (2013)]{Mortier-2013} Mortier, A., Santos, N. C., Sousa, S. G., et al.\ 2013, \aap, 557, A70
\bibitem[Press et al. (1992)]{Press-1992} Press, W.~H., Teukolsky, S.~A., Vetterling, W.~T., \& Flannery, B.~P.\ 1992, Cambridge: University Press, 1992, 2nd ed.,
\bibitem[Saffe (2011)]{Saffe-2011} Saffe, C., Revista Mexicana de Astronomía y Astrofísica Vol. 47, pp. 3-14 (2011)
\bibitem[Santos (2004)]{Santos-2004} Santos, N. C.; Israelian, G.; Mayor, M., \aap, 415, 1153
\bibitem[Sneden (1973)]{Sneden-1973} Sneden, C.~A.\ 1973, Ph.D.~Thesis,
\bibitem[Sousa et al. (2007)]{Sousa-2007} Sousa, S.~G., Santos, N.~C., Israelian, G., Mayor, M., \& Monteiro, M.~J.~P.~F.~G.\ 2007, \aap, 469, 783
\bibitem[Sousa et al. (2008)]{Sousa-2008} Sousa, S.~G.,  Santos, N.~C., Mayor, M., et al.\ 2008, \aap, 487, 373
\bibitem[Sousa et al. (2011)]{Sousa-2011a} Sousa, S.~G.,  Santos, N.~C., Israelian, G., et al.\ 2011, \aap, 526, id.A99
\bibitem[Sousa et al. (2011)]{Sousa-2011b} Sousa, S.~G.,  Santos, N.~C., Israelian, G., et al.\ 2011, \aap, 533, id.A141
\bibitem[Stetson et al. (2008)]{Stetson-2008} Stetson, P. B. \& Pancino, E. 2008, PASP, 120, 1332
\bibitem[Torres et al. (2012)]{Torres-2012} Torres, G., Fischer, D. A., Sozzetti, A., et al.\ 2012, APJ, 757, 2, 161, 14
\bibitem[Tsantaki et al. (2013)]{Tsantaki-2013} Tsantaki, M., Sousa, S. G., Adibekyan, V. Zh., et al.\ 2013, \aap, 555, A150

\end{thebibliography}
